# Using MRI Cell Tracking to Monitor Immune Cell Recruitment in Response to a Peptide-Based Cancer Vaccine

Marie-Laurence Tremblay[1], Christa Davis[1], Chris V. Bowen[1,2,3], Olivia Stanley[1], Cathryn Parsons[1], Genevieve Weir[5], Mohan Karkada[6], Marianne M. Stanford[5,7], *Kimberly D. Brewer[1,2,3,4,7]

[1]Biomedical Translational Imaging Centre (BIOTIC), Halifax, NS, Canada

[2]Department of Diagnostic Radiology, Dalhousie University, Halifax, NS, Canada

[3]Department of Physics and Atmospheric Science, Dalhousie University, Halifax, NS, Canada

[4]School of Biomedical Engineering, Dalhousie University, Halifax, NS, Canada

[5]Immunovaccine Inc., Halifax, NS, Canada

[6]Wyss Institute at Harvard Medical School, Boston, MA, USA

[7]Department of Microbiology and Immunology, Dalhousie University, Halifax, NS, Canada

Correspondence should be addressed to KDB (brewerk@dal.ca)

**Abstract**

**Purpose:** MRI cell tracking can be used to monitor immune cells involved in the immunotherapy response, providing insight into the mechanism of action, temporal progression of tumour growth and individual potency of therapies. To evaluate whether MRI could be used to track immune cell populations in response to immunotherapy, $CD8^+$ cytotoxic T cells (CTLs), $CD4^+CD25^+FoxP3^+$ regulatory T cells (Tregs) and myeloid derived suppressor cells (MDSCs) were labelled with superparamagnetic iron oxide (SPIO) particles.

**Methods:** SPIO-labelled cells were injected into mice (one cell type/mouse) implanted with an HPV-based cervical cancer model. Half of these mice were also vaccinated with DepoVax$^{TM}$, a lipid-based vaccine platform that was developed to enhance the potency of peptide-based vaccines.

**Results:** MRI visualization of CTLs, Tregs and MDSCs was apparent 24 hours post-injection, with hypointensities due to iron labelled cells clearing approximately 72 hours post-injection. Vaccination resulted in increased recruitment of CTLs and decreased recruitment of MDSCs and Tregs to the tumour. We also found that MDSC and Treg recruitment was positively correlated with final tumour volume.

**Conclusion:** This type of analysis can be used to non-invasively study changes in immune cell recruitment in individual mice over time, potentially allowing improved application and combination of immunotherapies.

**Introduction**

Immunotherapies are one of the fastest growing classes of cancer therapies with the potential to rival traditional therapies as frontline cancer treatments (1-3). Their rapid adoption has been driven by monoclonal antibodies (4-6), checkpoint inhibitors (7-11) and/or cytokine treatments capable of eradicating tumours through manipulation of the immune system. Cancer vaccines (12-14) are another type of T cell based immunotherapy capable of activating the immune system that are currently in development and undergoing clinical testing.

Cancer progression is routinely monitored using imaging to assess traditional clinical outcomes such as tumour size. However, molecular imaging can also be a valuable research tool to explore and understand longitudinal molecular mechanisms initiated by novel cancer therapies both preclinically and clinically (15-22). A variety of imaging modalities (i.e. magnetic resonance imaging (MRI), positron emission tomography (PET), or fluorescence) have been used to investigate the effect of these therapies pre-clinically (15-22). Fluorescence or bioluminescence optical imaging is common in the context of monitoring intensity changes and migration of therapeutics conjugated to fluorescent tags or specific cell types genetically modified for bioluminescence. Both are used as reliable pre-clinical markers of interest (23-25). However, optical imaging has limited tissue penetration depth and spatial resolution and lacks internal anatomy information provided by other imaging modalities such as MRI and PET/computed tomography (CT).

MRI has been extensively used preclinically to explore immunotherapy behaviour by tagging individual vaccine components with superparamagnetic iron oxide (SPIO), permitting longitudinal tracking of components over time (22). MRI imaging can also be used to track specific cell populations labelled with SPIO (26-31). Previous research using MRI has achieved *in vivo* detection and visualization of single cells labelled with SPIO (32,33)*,* which has been used to follow metastases originating from single cells visible by MRI (34). MRI has also been used to track the migration of cells injected for direct therapeutic effect for various diseases, including dendritic and cytotoxic T cell injections used as cancer therapies (30,31,35-37). However, rather than evaluate cellular migration patterns of a cellular therapy, we are tracking immune cells involved in the cancer immune response, providing insight on the mechanism of action, temporal progression and individual potency of immunotherapies.

Cytotoxic $CD8^+$ T cells (CTLs) are key cells in the fight against cancer (38). However, research has demonstrated that CTLs deployed directly as the sole cancer therapy have negligible therapeutic anti-tumour response for solid tumours (39). Immune suppressor cells, such as regulatory T cells (Tregs) and myeloid derived suppressor cells (MDSCs), are recruited by tumours to inhibit the cytotoxic effects of CTLs (40-44). The increased presence of Tregs and other suppressor cells are primarily responsible for dampening the vaccine-induced immune responses that target antigens over-expressed by cancer cells.

The success of cancer immunotherapies depends heavily on the interplay between cytotoxic and suppressive cell populations (41). Monitoring the location and migration of these cells during tumour growth and/or following therapy can be critical for a successful immunotherapy approach. It facilitates an understanding of the mechanism of action as well as efficient assessment of dose levels, booster timing, and the introduction and interaction of additional treatments. The present study aims to achieve this using modern cell tracking with SPIO and MRI methods. The migration of CTLs has been extensively studied during cytotoxic T cell therapies (36,45,46), but limited information is available on the migration of MDSCs and Tregs in cancer models. Furthermore, cellular migration following an active therapy, such as vaccination, has not been reported.

In this study, MRI cell tracking was used to monitor the migration and recruitment of SPIO-labelled CTLs, Tregs and MDSCs in a HPV16 expressing tumour mouse model (C3: (47,48) following immunotherapy. Cells were tracked in mice that were untreated or vaccinated with a peptide-based cancer vaccine in a DepoVax$^{TM}$ formulation (12). DepoVax$^{TM}$ (DPX) is a novel immunotherapy formulation designed to elicit sustained immune responses (12,43,49,50) via prolonged antigen release (21) and proprietary adjuvants. The migration of CTLs, Tregs and MDSCs were evaluated in response to either the presence or absence of DepoVax immunotherapy at the group level and on an individual basis.

**Methods**

***Reagents*** – Sources for all reagents, antibodies, and media recipes can be found in supporting materials and Supporting Table S1.

***Cancer Cell Line.*** The murine cervical cancer C3 cell line (47,48) (cryopreserved in fetal bovine serum (FBS) +10% DMSO in liquid nitrogen) were obtained from Dr. Martin Kast from the University of Southern California. C3 cells were thawed at 37°C, the freezing media was removed, and cells were cultured in C3 media at 37°C in a standard incubator with 5% $CO_2$.

***Vaccine Formulation.*** All peptides were synthesized by PolyPeptide Group with >90% purity. The CD8 epitope HPV16 $E7_{49-57}$ (RAHYNIVTF; R9F) and the universal T helper peptide $TT_{947-967}$(FNNFTVSFWLRVPKVSASHLE; F21E), were used in vaccine formulations and were prepared as a proprietary DPX formulation (43), referred to herein as DPX-R9F. Briefly, a lipid-mixture containing phosphatidyl choline and cholesterol in a 10:1 ratio (w:w) (Lipoid GmBH), R9F (5 μg/dose), F21E (5 μg/dose), and a proprietary polynucleotide based adjuvant (20 μg/dose) were formulated in 40% tert-butanol, lyophilized and resuspended in Montanide ISA 51 VG (SEPPIC). R9F is the epitope specific to the C3 cancer model.

***Tumour Challenge and Vaccination***. C57BL/6 female mice (4-6 weeks old, Charles River Laboratories) were housed with food and water *ad libitum* under filter top conditions. C57BL/6 mice were used as both donor and recipient mice for all MRI experiments. Experiments involving mice were carried out in accordance with ethics protocols approved by the University Committee on Laboratory Animals at Dalhousie University, Halifax, N.S., Canada.

All mice were implanted with $5 \times 10^5$ C3 cells subcutaneously (s.c.) into the left flank (study day 0). Mice were either i) untreated (n=13), or ii) vaccinated with DPX-R9F (50 μL) 15 days post-implantation (n=13). Vaccines were delivered via a single s.c. contralateral immunization (right flank). Tumour sizes were measured with calipers on a weekly basis for the first 3 weeks, then bi-weekly using the following formula: [longest measurement × (shortest measurement)$^2$] ÷ 2. Mice were scanned 24 and 48 and 72 hours post-injection and terminated approximately 96 hours post-cell injection. Final tumour volumes were determined post mortem. Lymph nodes (LNs), spleen and tumour were harvested and frozen in OCT:20% sucrose (2:1, v/v) for immunohistochemistry.

***CTL Isolation and Labelling***. One week prior to cell injections, inguinal, axial, brachial and mesenteric LNs were isolated from donor mice that also had tumours and had the same treatment as the recipient mice (either control or vaccine). CTLs were isolated from LNs via panning. CTLs were enriched by incubation with rat anti-mouse CD4 (25 μg/mL) and rat anti-mouse IgG1κ (50 μg/mL) at a density of $1 \times 10^8$ cells/mL. Lymphocytes were then incubated on a petri dish coated with goat anti-rat IgG (10 μg/mL). The enriched CTL population was cultured *in vitro* at a density of $5 \times 10^5$ cells/mL in complete RPMI media (cRPMI) supplemented with mouse CD28 (1 μg/mL), gentamycin (5 μg/mL), IL-2 (20 U/mL), and IL-12 (100 ng/mL) in a coated CD3 (2 μg/mL) culture flask to encourage cellular proliferation and activation. Cells were monitored over five days and kept at a density of $5\text{-}10 \times 10^5$ cells/mL. Fresh cRPMI with IL-2 (20 U/mL) was added as nutrients were depleted. Two days after CTL isolation, splenocytes were isolated from disease- and treatment-matched mice to act as antigen presenting cells (APCs) to CTLs, stimulated with LPS (10 μg/mL) and cultured *in vitro* in C3 media for two days. Splenocytes were then treated with mitomycin-c (50 μg/mL) for 20 min, washed thoroughly, added to the CTLs culture at a ratio of 1:6 APC:CTLs and incubated with R9F (10 μg/mL) to encourage further priming. At day 7, all cells were washed and iron loaded by incubation at a density of 3-4 million cells/mL in cRPMI supplemented with IL-2 (100 U/mL) and SPIO (0.1 mg/mL; 30 nm, Rhodamine B Molday ION, Biopal) for 20-24h.

***MDSC isolation and labelling***. One day before cell injections, spleens were isolated from donor mice that also had tumours and had the same treatment as the recipient mice (either control or vaccine). MDSCs were purified from other splenocytes using a magnetic positive selection kit with anti-Gr-1 biotinylated antibody and a biotin specific MACS kit (Miltenyi-Biotec). After isolation, cells were incubated overnight in cRPMI with 0.1 mg/mL of SPIO.

***Treg cell isolation and labelling***. Six days prior to cell injections, inguinal, axial, brachial and mesenteric LNs were isolated from donor mice that also had tumours and had the same treatment as the recipient mice (either control or vaccine). CD4+ T cells were purified from other lymphocytes via positive CD4+ magnetic selection kit (EasySep, Stemcell technologies). CD4+ T cells were differentiated into Tregs by incubation for 5 days in cRPMI supplemented with IL-2 (500 U/mL), CD28 (1 μg/mL), gentamycin (5 μg/mL), TGF-β (5ng/mL; EMD Millipore), and rapamycin (100 nM) in a culture flasks coated with 2 μg/mL CD3. At day 5, cells were washed and iron loaded by incubation at a density of 3-4 million cells/mL in cRPMI supplemented with IL-2 (500 U/mL), TGF-β (5 ng/mL), rapamycin (100 nM) and SPIO (0.1 mg/mL) for 20-24h.

***Cell injection preparation***. SPIO-labelled cells (CTLs, Tregs or MDSCs) were washed thoroughly and resuspended in HBSS$^{++}$ with 20 mM HEPES buffer for 200 μL injection of either 10 million CTLs (n=10, treated: n=5) or 5 million MDSCs (n=8, treated: n=4) or 4 million Tregs (n=8, treated: n=4). All cells were injected intravenously into the tail vein on day 28. Numbers of injected cells were chosen based on preliminary MRI qualitative data. Cells not used for injection were used for a trypan blue viability test and analysed by spectrophotometry using a Prussian blue assay (51) to assess iron loading by UV/VIS absorbance spectrophotometry ($\lambda$ = 620 nm). Cellular iron was quantified by lysing cells overnight in 100 μL of 1M HCl. $K_4Fe(CN)_6$ (1N, 100 μL) was then added, resulting in an intense blue colour in the presence of iron. The resulting absorbance was compared to a calibration curve of known SPIO concentrations.

***Cell culture purity assessment by FACS***. Purity analysis of the *in vitro* culture were assessed by flow cytometry using cell samples from respective cultures (~$2 \times 10^6$ cells) prior to SPIO uptake. All fluorescent antibodies were anti-mouse. CTLs were stained with CD8α (CD8)-efluor660, CD3-peridinin-chlorophyll proteins efluor 710 (PerCP), CD4-phycoerythrin (PE), and CD11c-PE. Treg cells were labelled with CD25-PE, FoxP3-allophycocyanin (APC), and CD4-PerCP. MDSCs were labelled with CD11b-PE and Ly-6G (Gr-1)-efluor660. Before staining, cells were pre-incubated for 10 min in Fc block (rat anti-mouse CD16/CD32, clone 2.4G2) then incubated for 30 minutes at 4°C with respective surface markers. Cells were fixed using 2% PFA in PBS. Intracellular FoxP3 staining was performed by treating cells with 1X permeabilizing buffer (Saponin; eBioscience) and staining with FoxP3-APC for 2 hours at 4°C. OneComp ebeads (eBioscience) were used for isotype controls. Cells were analysed using a two laser FACSCalibur equipped with CellQuest Pro software (BD Biosciences) and set to collect $5 \times 10^4$ events. Analysis was performed using FCS Express 6 flow software (De Novo Software) and all samples were gated for live cells. $CD4^+CD25^+FoxP3^+$ Tregs cells were further gated for $CD4^+CD25^+$ cells then $FoxP3^+$ to properly establish Tregs.

*Cell suppression assays*. Transgenic C57BL/6 UbC-GFP mice were used for Treg cell isolation and culture (described above). Responder $CD4^+$ T cells (Tresp) were isolated from LNs from C57BL/6 mice by $CD4^+$ positive selection as described above for Treg cell culture. Tresp cells were labelled with the cell proliferation dye efluor670 (0.5 μM; eBioscience) following directions by the manufacturer. Tresps were co-cultured at $1 \times 10^5$ cells/well in a 96-well plate coated with CD3 with varying ratios of Tregs (Tresp:Tregs = 1:1, 1:5, and 1:10) in cRPMI supplemented with 1 μg/mL of CD28. Cells were incubated at 37°C, 5% $CO_2$, for 72 hours. After harvest, proliferation was determined by flow cytometry (as described above) and gated for $GFP^-$ $efluor670^+$ cells. Treg controls were also added to the analysis ensuring that the $CD4^+CD25^+FoxP3^+$ Treg population was >80% at day 0. The suppressive ability of Tregs was assessed by comparison of Tresp at day 0 to Tresp at day 3. The suppressive ability of MDSCs isolated using this methodology was previously assessed in (43).

*Data acquisition and MR Imaging*. All data were acquired on a 3T magnet equipped with a 21 cm i.d. gradient coil (200mT/m; Magnex Scientific, Oxford, UK) interfaced with a Varian DD Console (Varian Inc., Palo Alto, CA, USA). A 30 mm i.d. quadrature transmit/receive RF coil (Doty Scientific, Columbia, SC, USA), was used to image tumours, vaccination sites, and inguinal LNs simultaneously.

Anatomical images were obtained using a 3D balanced steady-state free precession (bSSFP) (52) sequence (repetition time (TR)/echo time (TE) = 8/4 ms, flip angle = 30°, a 38.4×25.5×25.5 mm field of view (FOV), 256×170×170 matrix centered on the torso, 150 μm isotropic resolution). Four signal averages were acquired with four separate RF phase increments to displace and remove off-resonance banding artifacts (52) for a total scan time of approximately 64 minutes per animal. Mice received a baseline scan prior to SPIO-labelled cell injection on study day 27, 24 hours prior to cell injection. Mice were scanned 24, 48 and 72 hours post-cell injection.

***MRI Image Analysis and Statistical Analysis***. Volumetric segmentation was performed by a single observer, and confirmed by a second independent reviewer. All images were first zero-padded (interpolated to a higher resolution grid to increase the effective resolution and image quality) using ImageJ (NIH). Images were analysed in RView for each mouse (53,54). A semi-automated region growing algorithm (54) was implemented to perform individual 3D segmentations to determine tumour volumes, left inguinal LN (LLN, tumour draining) and right inguinal LN volumes (RLN, vaccine draining).

Tumour and lymph node regions of interest (ROIs) were hand-drawn in VivoQuant (Invicro) and verified by a separate reviewer. At each time point, the 3D tumour and lymph node ROIs encompassed the entire tissue structure of interest (i.e. tumour ROIs consisted of the entire tumour) to minimize difference due to positional changes. ROI histograms were extracted and cell recruitment was assessed by integrating point estimates of iron concentration in each voxel within tumour ROIs through all voxels in the tumour that exhibit MRI signal changes upon cell injection. This semi-quantitative approach (22) estimates the ROI iron mass (M[Fe]) for inhomogeneous iron distributions within the tumour, by weighting each voxel per its negative log intensity ratio (a proportionate estimate of iron concentration):

$$M[Fe] = -\int Ln\left(\frac{I}{I_{median}}\right) \times \left[\frac{Voxel_{bin}}{Voxel_{total}}\right] \qquad (1)$$

The natural log of each histogram bin's signal value (*I*) is divided by the median signal value of the whole tumour ROI ($I_{median}$). This value is multiplied by the normalized bin size, i.e. number of voxels in that bin ($Voxel_{bin}$), divided by total number of voxels within the tumour ROI ($Voxel_{total}$), resulting in a relative mass of iron M[Fe] for each bin. For lymph nodes, we used the M[Fe] integrated over the left side of the intensity distribution (i.e. all intensities less than the median intensity), hereafter represented as M[Fe]$_{median}$. For tumours, due to the increased heterogeneity, and to increase the likelihood that only voxels with iron-labelled cells are considered, M[Fe]s were summed over only the lowest 10% of the intensity distribution. This value is hereafter represented as M[Fe]$_{10}$. Use of M[Fe]$_{10}$ for lymph nodes was not practical due to their small size and fewer voxels. All M[Fe] values were then compared to baseline (i.e. pre-iron injection) to determine the % increase in M[Fe] at 24 and 48 hours post-injection.

Statistical correlation analyses were done in Prism 6 (GraphPad Software) post-hoc. SPSS (IBM analytics) was used to do a mixed ANOVA to evaluate the M[Fe]$_{10}$ for each cell type over time and to evaluate the change in LN intensities. One-way ANOVAs were used to compare M[Fe]$_{10}$ for untreated vs vaccinated mice at 24 hours post-cell injection.

***Immunohistochemistry***. All mice used for immunohistochemistry were terminated at 24 hours post-injection (separate group of mice from MRI). Transgenic C57BL/6 UbC-GFP mice were used as donor mice and normal C57BL/6 mice were used as recipients with tumour implants and vaccination done as described previously. Cryosectioning of tissues was performed at -20°C with 5 μm thickness. An hour after sectioning, the tissues were rehydrated in PBS and fixed in 4% PFA for 15 minutes. Tissues were washed 3x in PBS and incubated in block (0.1% BSA in PBS) for 30 minutes. Tissues of mice that received either MDSC or CTL cell injections were incubated with either Ly-6G (Gr-1)-efluor660 or CD8α-efluor660, respectively, at 1:200 in block for 1 hour at 4°C. Tissues of mice that received Treg cells injections were first permeabilized in 1X permeabilizing solution for 10 min, and then incubated in FoxP3-APC at 1:100 in block for 1 hour. After staining, slides were washed 3x in PBS and mounted with Fluoromount-G containing DAPI (eBioscience).

***Fluorescence confocal microscopy***. Tissues were visualized using a Zeiss LSM 710 laser scanning confocal microscope (Carl Zeiss SBE, LLC) equipped with an XBO 50W lamp for DAPI fluorescence ($\lambda_{ex}$ 365, $\lambda_{em}$ 420 nm), an Argon laser for GFP fluorescence ($\lambda_{ex}$ 488 nm) equipped with a 515-565 band pass filter, and a HeNe laser for 548 and 633 nm excitation equipped with a low pass filter at 590 nm. All images were acquired using 4 averages per pixel at 1-1.5 A.U., a pixel dwell time of 6.30 μsec, 1024×1024 resolution, on 0.9-1.2 µm slices. Nuclei were stained with DAPI (a stain that binds strongly to DNA) were used to visualize tissue morphology. Identification of cells within tumours from adoptive transfers was determined by overlay of GFP, Rhodamine B and the cell type identification (Gr-1-efluor660 for MDSCs, CD8α-efluor660 for CTLs or FoxP3-APC for Tregs) signals. Images were visualized and processed using FIJI (ImageJ, NIH) (55).

**Results**

*Cell culture, labelling and purity.* Cell purity was assessed by FACS before incubation with SPIO to ensure >80% purity. CTLs were >80% pure (Supporting Fig. S1A), with ~11% of the cell population identified as APCs (CD11c$^+$) used for cell activation during *in vitro* expansion. The suppressive cell populations CD4$^+$CD25$^+$FoxP3$^+$ Tregs and Gr-1$^+$CD11b$^+$ MDSCs were at >80% purity after cell isolation and culture (Supporting Fig. S1B and C). Cell suppression assays were performed to determine the suppressive nature of both the MDSCs and Tregs (Supporting Fig. S2 for Tregs, see (43) for previous evaluation of the MDSCs used in this experiment – cells for both experiments were prepared and evaluated at the same time in 2014). Both cell types were successfully able to suppress cellular proliferation of responder T cells. Labelling cells with SPIO did not significantly affect cellular viability as assessed by trypan blue before cell injection. Iron loading levels, determined by the Prussian blue iron assay, were between 3-5 pg iron/cell for CTLs, between 5-7pg iron/cell for Tregs and approximately 30 pg iron/cell for MDSCs.

*Tumour growth.* For this study we chose to administer the immunotherapy at a later time point than previous cancer studies with DPX-R9F (22,56,57) such that there would be some response from the therapy, but not complete tumour remission. As seen in Fig. 1A, vaccinating at 15 days post-implant resulted in tumours that were statistically significantly smaller than untreated controls on study day 31 (p<0.001, two-tailed student t-test). Mice injected with MDSCs had significantly larger tumours than mice injected with Tregs (p<0.05, multiple comparison t-tests), but not significantly larger than mice injected with CTLs (Fig. 2B) in the vaccinated group. It is possible there may have been some difference in vaccine efficacy driving decreased tumour volumes in mice receiving MDSCs vs Tregs, as experiments were done at different times and vaccinated from different vaccine batches.

There was no significant difference in tumour volumes in vaccinated mice that received CTLs compared to Tregs. In untreated mice, there were no significant differences in tumour volumes regardless of cell type that was injected.

*Immunohistochemistry*. To differentiate between injected cells and innate/cancer cells, C57BL/6 UbC-GFP transgenic mice were used as donor mice and cells isolated from donor mice were injected I.V. into non-transgenic C57BL/6 diseased matched mice. All protocols were performed the same otherwise. The mice receiving cell injections were terminated 24 hours post-cell injection to maximize $GFP^+$ and SPIO-rhodamine B fluorescence. Validation of recruitment of SPIO-labelled cells to tumours was performed by co-localizing rhodamine B fluorescence from the SPIO nanoparticles with $GFP^+$ signals from the injected CTLs, Tregs or MDSCs cells (Fig. 2). As a third verification, cell specific fluorescent antigens were used to label either CTLs ($CD8^+$), Tregs (FoxP3), or MDSCs (Gr-1). As expected, both GFP and rhodamine B fluorescence perfectly co-localize for each cell type, indicating adoptively transferred cells injected I.V. did localize to tumours, and can be located amongst surrounding tumour cells.

*Qualitative Descriptions of MR images*. All tumours were implanted at the same location, however due to individual growth variability within mice, tumours may appear to be placed slightly different. Additionally, images in Figs. 3, 4 & 5, have been zoomed in to focus on tumours. Localized regions of hypointensities were observed approximately 24 hours after CTL injections, indicating the presence of SPIO-labelled cells in the tumours of both untreated and vaccinated mice (Fig. 3). In untreated mice, cells were mostly concentrated at the tumour periphery. However, there were instances of cells localizing further centrally within the tumour (Fig 3; arrows). In vaccinated mice, clusters of hypointensities were located both at the tumour periphery and in central regions. At 48 hours post-cell injections, hypointense regions disappeared from the tumour, which indicated that either cells are clearing from the tumour or that SPIO nanoparticles within the cells are undergoing degradation via the lysosomal pathway (58). CTL recruitment to the LNs was visibly apparent 24h post cell injection (Fig. 6).

Suppressive SPIO-labelled MDSCs preferentially localize throughout the tumour in untreated mice as indicated by strong hypointense regions 24h post cell injections (Fig. 4). MDSCs remain visible for 48h post-injection and iron generally begins to clear 72 hours post-injection. In contrast to the untreated group, three of four vaccinated mice lacked hypointensities, implying that recruitment is scarce within the tumour microenvironment in this group. Only one vaccinated mouse in the group displayed hypointensities demarcating cells. There was no observed MDSC recruitment to LNs in both treated and untreated mice, reflected by no large change in signal intensity from images acquired before/after MDSC injection. This was not unexpected since MDSCs preferentially migrate to the tumour microenvironment (59).

As with MDSCs, Tregs were recruited to the tumour. Recruitment was visualized as hypointense regions throughout the tumour 24h post-cell injection in untreated mice (Fig. 5) and iron started clearing by 48h. Minimal hypointensities were detected in vaccinated mice throughout the tumour both 24 and 48 hours post cell injection, indicating minimal recruitment. Noteworthy, one of the four vaccinated mice injected with SPIO-labelled Tregs underwent complete remission with no detectable tumour at day 27; therefore, mapping the migratory patterns of Tregs for this mouse was not possible. As with CTLs, Tregs were abundantly recruited to lymph nodes and the iron persisted up to 144 hours (Fig. 6).

***Semi-quantitative analysis of SPIO-labelled cells in LN***. Signal intensity distributions were analysed in the form of histograms and used to semi-quantitatively analyse the recruitment of CTLs, Tregs, and MDSCs at 24 and 48 hours post cell-injection compared to baseline. Only voxels within lymph nodes having signal intensities less than the median were used for quantitation ($M[Fe]_{median}$; equation 1) since it was deemed a more accurate representation of SPIO-labelled cell number and best characterized the MR image, particularly given the small size of the lymph nodes which prevented the use of a smaller portion of the distribution. The vaccine-draining lymph is the right inguinal lymph node, whereas the tumour draining lymph node is the left inguinal lymph node. Relative to unvaccinated mice, vaccinated mice showed trends of increased mean recruitment of CTLs and decreased mean recruitment of Tregs in both the tumour-draining and vaccine draining inguinal LNs (Fig. 7; images in Fig. 6). Although the means were different, they were not statistically significant. There were no quantifiable signal changes in LNs for mice receiving MDSC injections.

***Semi-quantitative analysis of SPIO-labelled cells in tumours***. Tumour mean signal intensity is not reflective of total iron content due to its continuously changing heterogeneous environment and therefore intensity histograms of tumour ROIs were extracted and used to filter out high intensity features (i.e. edema). As with lymph nodes, signal intensity distributions within the tumour were analysed in the form of histograms. Due to a larger number of voxels within tumours, those voxels having the lowest 10% of signal intensities were used for quantitation ($M[Fe]_{10}$; equation 1) and results at 24 and 48 hours were again compared to $M[Fe]^{10}$ values at baseline.

Vaccinated mice injected with CTLs had an overall higher $M[Fe]_{10}$ than untreated mice (Fig. 8E) using our semi-quantitative analysis. Although the means seem to visually differ between treated and untreated mice, the large individual variability makes it statistically insignificant. However, the mean Treg $M[Fe]_{10}$ value in vaccinated mice was significantly lower than untreated mice by one-way ANOVA at 24 hours post cell-injection ($p$-value < 0.05) (Fig. 8C). The mean MDSC $M[Fe]_{10}$ in untreated mice was statistically significant at 48h ($p$-value < 0.1) but not 24h ($p$-value > 0.1) post cell injection by one-way ANOVA (Fig. 8A). We have generally found that iron signal in tumours from MDSC recruitment seems to peak at 48 hours post-cell injections, whereas for Tregs and CD8s, the iron signal peaks at 24 hours. This is likely due to differences in cellular infiltration and division rates between different cell types (MDSCs are fully differentiated cells and are not likely to divide and proliferate *in vivo*).

Individual mouse correlations were made between tumour size and $M[Fe]_{10}$ for MDCSs, Tregs and CTLs (Fig. 8B, D, E). SPIO-labelled MDSC $M[Fe]_{10}$ showed a statistically significant positive correlation with final tumour volume ($r$ = 0.8226, $p$-value < 0.01) (Fig. 8B) 24h post-injection. Recruitment of SPIO-labelled Tregs showed a statistically significant positive correlation with final tumour volume ($r$ = 0.6113, p-value < 0.1) at the 24h post-injection time point (Fig. 8D). This implies an increase in recruitment of MDSCs and Tregs (as measured by $M[Fe]_{10}$) with increasing tumour size for both untreated and vaccinated mice. CTL recruitment was very weakly negatively correlated with increasing tumour size (one outlier was removed according to Grubb's statistical test (60) in Prism 6 (Fig. 8F)), but there was no statistical significance.

**Discussion**

This work demonstrates the possibility to monitor and measure semi-quantitative changes in the migration of $CD8^+$ cytotoxic T cells, regulatory T cells and myeloid derived suppressor cells in tumour bearing mice following treatment with the vaccine based immunotherapy DepoVax™ (DPX). The presented technique is a potentially valuable tool for gaining a better understanding of underlying biological mechanisms *in vivo* in response to immunotherapies in preclinical studies. Isolated SPIO-labelled cells cultured *in vitro* were visualized by MRI as early as 24h post-cell injection (Figs. 3 to 6) and iron was cleared approximately 72h post-injection. Vaccination with the peptide-based vaccine DPX-R9F, as seen by the semi-quantitative metric for iron mass, $M[Fe]_{10}$, caused a decrease in MDSC and Treg recruitment and a small increase in CTL recruitment in tumours (Fig. 8), albeit one that was non-significant. CTLs were recruited to both tumour and vaccine draining inguinal LNs, particularly for vaccinated mice (Fig. 7A and B). However the difference between vaccinated and control mice was not significant. Treg recruitment decreased to both inguinal LNs with vaccination (Fig. 7C and D). These data are consistent with previously published data using the cervical tumour model (12,43,61). We have clearly demonstrated that vaccination with the DPX-R9F formulation decreases the recruitment of suppressive cells types (MDSCs and Tregs) to the tumour and mildly increases the immune system's first line of defence: CTLs.

MRI immune cell tracking is a powerful technique for monitoring treatment efficacy due to its sensitivity in detecting individual variations in response to therapy. An increase in MDSC and Treg recruitment was evident 24h post-injection between vaccinated and untreated mice but also at an individual level, positively correlating tumour size to cellular recruitment within the tumour regardless of treatment (Fig. 8A, C). These results were expected given that larger, more advanced stage tumours are more effective in recruiting suppressive immune cells (62). This is likely one of the main reasons why vaccinated mice have significantly fewer Tregs and MDSCs. Additionally, previously published work (12,61) has described that vaccination with DPX caused a significant decrease in MDSCs and Tregs in the spleen and blood of tumor bearing mice (measured using FACS) as well as within tumors (via immunohistochemistry).

The difference in CTL recruitment to the tumour was minimal for vaccinated mice compared to untreated mice at both time points (Fig. 8E). The mean increase observed in Fig. 8E was strongly influenced by one mouse and recruitment to the tumour did not appear to depend on tumour size. There were large variations within vaccinated mice; some had barely detectable CTL cell recruitment, some exhibited small amounts of CTL recruitment (15-20% increase in $M[Fe]_{10}$), and one mouse had an extremely large increase in $M[Fe]_{10}$ (147%). This observed variability is not uncommon given that a hallmark of immunotherapy is individual variation (63). This sensitivity to individual-level variability may be beneficial for determining and predicting individual responses to therapy. It is also possible that the chosen 4 weeks post-implant time point may not have been optimal for monitoring CTL migration to the tumour, particularly in response to treatment.

Vaccination resulted in a large increase in CTL recruitment to both inguinal LNs. Recruitment to the vaccine-draining LN (Fig. 7A) was expected due to the vaccine-driven immune response (22). In this response, the antigen presenting cells bring antigens from the vaccination site to the LNs for presentation to T cells like CTLs (22). Increased recruitment in the tumour-draining LN (Fig. 7B) likely indicates an active immune response to tumour cells or tumour cell antigens that may already be present in the tumour-draining LN.

Interestingly, although vaccination did cause a small increase in CTL recruitment to tumours and LNs, it was not the significant increase that might be expected. A potential explanation for this is likely related to the functionality of these CTLs. Prior to injection, these cells have been activated and primed to be cytotoxic effector CD8+ T cells. However, once they are injected, they are subject to the same environment as host cells, particularly the extremely suppressive tumour microenvironment. In this case, CTLs in the vaccine group may be more successful at retaining their functionality as cytotoxic cells, whereas CTLs injected into control mice become dysfunctional. Previous work (64) demonstrated that genes linked to clonality and cytotoxic activity were upregulated in cytotoxic T cells in response to therapy, but actual cell numbers remained unchanged. As indicated by the decreased recruitment of suppressive cells (both Tregs and MDSCs) in vaccinated mice, the tumour microenvironment will likely affect functionality differently in each group, which is unfortunately not detectable with MRI cell tracking. Additional antigen presentation will also occur in vaccinated mice due to the presence of sustained antigen release. Previous publications (12, 61) have also speculated that it is the interplay of regulatory and effector T cells which will drive success vaccine responses, as opposed to solely CTL activity alone.

Although the use of MRI cell tracking with SPIO has a tremendous amount of potential for studying immune responses to therapy, there are currently some important limitations. Hypointensities associated with iron oxide are not specific to SPIO. Tumours are notoriously heterogeneous: necrosis is a source of negative contrast and edema is a source of positive contrast with this pulse sequence. Attempts to compensate for these discrepancies were made by ensuring that hypointensities of interest were not present in baseline images and also cleared by three days post-injection. Contrary to SPIO-labelled cells, necrotic regions remain hypointense and persist past the 72h time point observed in this study. However, for longitudinal studies, adding additional baseline and follow-up scans is both time and resource-intensive. Using the lowest 10% of the intensity distribution in the semi-quantitative technique herein as representative of iron concentration is empirically reasonable but still somewhat arbitrary. Additionally, for later timepoints, particularly 48 hours and 72 hours, it is unclear as to whether cells are actually 1) clearing from the tumour, 2) dividing, thereby diluting the visible iron, or 3) the iron is degraded within lysosomes and is no longer superparamagnetic. More quantitative imaging techniques are necessary to provide more accurate data, which would allow better understanding of individual-based differences, although there still would be limitations related to assessing functionality of cells necessitating the addition of histology and flow cytometry.

Specificity and quantitation issues can be overcome by the use of fluorinated cellular labelling agents (29,65-68). Previous studies have used fluorine cell labelling for several different cell types administered as an external cell-based therapy (i.e. dendritic cells (31,65), stem cells (67)) or to label macrophages to study inflammation (29). Fluorine labelled cells can be detected with precise specificity using specialized surface coils or dual-tuned hydrogen/fluorine coils. Fluorine-based MRI cell tracking can also yield quantitative information (29,67,68). Unfortunately, fluorine MRI is less sensitive than conventional proton MRI, making it difficult to detect smaller numbers of labelled cells, particularly at lower field strengths. It also requires more specialized equipment not necessarily available in all imaging facilities.

Quantitative information can also be obtained by using a pulse sequence like TurboSPI (69,70). Originally designed for imaging water in porous media that have long $R_2^*$ values but short $R_2$ values, Rioux et al. demonstrated that TurboSPI could also be applied to SPIO-labelled cells, which exhibited similar magnetic properties (69). Although TurboSPI traditionally requires long data acquisition times, the sequence was greatly accelerated for *in vivo* use by adding compressed sensing reconstruction (70). It is possible to extract $R_2^*$ maps from TurboSPI, which when combined with *in vitro* iron loading data, allows generation of cellular density maps. Although not specific to iron, $R_2^*$ maps allow the distinction between iron-labelled cells and other causes of tissue heterogeneity, such as necrosis, based on the voxel time courses (69).

We are currently working to implement TurboSPI and evaluate the use of $R_2^*$ and cellular density maps to monitor migration of CTLs and Tregs in response to several immunotherapies. It will be particularly important to research variations in individual responses to determine if the use of MRI as an immune cell tracking device can yield important information on individual responses to immunotherapy, and more importantly, provide insight into optimal therapy combinations.

Although this study was mainly designed as a proof of principle for tracking CTLs, Tregs and MDSCs in response to therapy, it will be particularly interesting to apply this technique to multiple time points, particularly to capture differences in cell migration in response to treatment over time. The longitudinal changes in immune behaviour in individual mice could potentially be linked with newer biomarkers such as LN swelling (56) and the immune-related response criteria (71) or iRECIST (72), as well as tumour volumes at the end of study.

**Conclusion**

This work demonstrates that MRI cell tracking can be used to monitor immune cell migration for various cell types in a cancer model in response to treatment with a peptide-based vaccine immunotherapy. MRI cell tracking has immense potential as a preclinical tool to increase our understanding of *in vivo* longitudinal dynamics of various immune populations in response to many immune therapies currently under development. Most importantly, it is possible to study changes in immune cell recruitment in individual mice over time, potentially allowing improved application and combination of therapies that can then be translated to the clinic.

**Acknowledgements**

The authors would like to acknowledge technical support from Alecia Mackay and Rajkannan Rajagopalan.

KDB received funding from the Nova Scotia Health Research Foundation and the IWK Health Centre. MLT received support from the Beatrice Hunter Cancer Research Institute Cancer Research Training Program (CRTP).

**Additional Information**

*Supporting material.* Supporting material accompanies this paper.

*Competing financial interests.* At the time of the study, KDB, GW, MK and MMS were employees of Immunovaccine Inc.

**Figure Captions**

**Figure 1** – *Final Tumour Volumes.* Final tumour volumes at study day 31 were measured with calipers. There was a statistically significant difference between untreated and vaccinated mice (unpaired student T-test, $p<0.0001$). There was no significant difference between untreated groups, regardless of cell type injected. For vaccinated mice, mice that received MDSC injections had significantly larger tumour volumes than mice that received Treg injections, but were not significantly larger than mice that received CTL injections.

**Figure 2** – *Immunohistochemistry of SPIO-Labelled Cells in Tumours.* D) Laser scanning confocal microscopy characterization of 5 μm C3 tumour tissues sections ~29 days post C3 implant and 1 day post cell injection. Injected cells were $GFP^+$ under the control of the ubiquitin C promoter (green) and labelled with SPIO rhodamine B nanoparticles (red). Tissues were labelled with a cell-specific antigen (yellow): Tregs = FoxP3-APC, CTLs = CD8a-efluor660, and MDSCs = Gr-1-efluor660. All tissues were embedded with Fluoromount-G containing DAPI (blue). The scale bars represent 25 μm.

**Figure 3** – *Representative bSSFP MR Images of tumours from Mice that Received SPIO-labelled $CD8^+$ cytotoxic T Cell Injection.* Top row is representative of untreated mice and bottom row of vaccinated mice. Images are from baseline (pre-injection), 24 hours, and 48 hours post-injection. In both untreated and vaccinated mice there is a large amount of CTL recruitment to both the periphery of the tumour and the tumour core (arrows) with hypointensities beginning to clear by 72 hours (data not shown). At baseline, mouse positioning causes the tumour to appear shifted posteriorly, however images are displaying the same central region for all days. L – left, R – right.

**Figure 4** – *Representative bSSFP MR Images of tumours from Mice that Received SPIO-Labelled MDSC Injection.* Top row is representative of untreated mice and bottom row of vaccinated mice. Images are from baseline (pre-injection), 24 hours, and 48 hours post-injection. In untreated mice there was a large amount of MDSC recruitment throughout the tumour (arrows), with hypointensities beginning to clear by 72 hours. MDSC recruitment was minimal in vaccinated mice. L – left, R – right.

**Figure 5 –** *Representative bSSFP MR Images of Tumours from Mice that Received SPIO-Labelled Treg Injection.* The top row is representative of untreated mice and the bottom row of vaccinated mice. Images are from baseline (pre-injection), 24 hours, and 48 hours post-injection. In untreated mice there is a large amount of Treg recruitment throughout the tumour (arrows), with some hypointensities clearing by 48 hours. Treg recruitment was decreased in vaccinated mice. L – left, R – right.

**Figure 6 -** *Representative bSSFP MR Images of LNs from Mice that Received SPIO-Labelled CTL or SPIO-Treg Injections.* Hypointensities in LNs represent CTL or Treg recruitment at 24h, 48h and 72h (CTLs) or 144h (Tregs). L – left, R – right.

**Figure 7 –** *Semi-Quantitative Assessment of T Cell Recruitment to Inguinal LNs.* Recruitment of either CTLs (A, B) or Tregs (C, D) to vaccine-draining inguinal LN (A, C) and tumour-draining inguinal LN (B, D) using the $M[Fe]_{median}$ as a metric. Vaccinated mice had increased recruitment of CTLs to both inguinal LNs and decreased recruitment of Tregs but differences were not statistically significant. Error bars represent standard error.

**Figure 8 -** *Semi-Quantitative Assessment of Cell Recruitment in Tumours and Correlation of Immune Cell Recruitment (Using $M[Fe]_{10}$) and Final Tumour Volume.*
Recruitment of MDSCs (A, B), Tregs (C, D), and CTLs (E, F) to the tumour using $M[Fe]_{10}$ as a metric. Vaccinated mice demonstrated decreased recruitment of MDSCs and Tregs to the tumour and increased recruitment of CTLs to the tumour. Significance judged using one-way ANOVA. Error bars represent standard error. Correlation of MDSC (B), Treg (D), or CTL (F) recruitment 24 hours post-injection with final tumour volume. Both MDSC and Treg recruitment ($M[Fe]_{10}$) was positively correlated with the final tumour volume. CTL recruitment ($M[Fe]_{10}$) was negatively correlated with final tumour volume but correlation was not significant. Dotted curved lines represent 95% confidence intervals. One outlier was removed from CTL correlation graph using Grubbs statistical outlier test. * represents $p<0.1$, ** $p<0.05$.

Supporting Figure S1. *Cell Purity of Isolated Cells.* Flow cytometry culture purity analysis for CTLs (A), Tregs (B) and MDSCs (C). Analysis was done immediately prior to iron labelling. Cells were gated for live cells and are representative of the full cell injection. CD11c control demonstrates that a small proportion (~10%) are APCs due to addition of APCs to culture for *in vitro* stimulation of T cells. FoxP3$^+$ cells were first gated for CD4$^+$CD25$^+$.

**Supporting Figure S2** - *Treg Supression Assay.* Treg-mediated suppression was measured using the cell proliferation dye efluor670 (eBioscience) at 0.5 μM. CD4$^+$ T cells (Tresp) were isolated from tumour-bearing C57BL/6 mice, incubated with efluor670, and activated using CD28 antibody on a plated CD3$^+$ 96-well plate. Tregs were isolated from transgenic GFP$^+$ mice. Tresps were cultured either alone (blue) or in the presence of Tregs at a 1:1 ratio (red). Cells were cultured for 72 hours and proliferation was determined by flow cytometry analysis gating for GFP$^-$ efluo670$^+$ cells. Tresp stimulation was confirmed by comparison to Tresp fixed at 0h of incubation (black).

**Supporting Table S1** - *Primary Antibodies and Sources.*

Figure 1

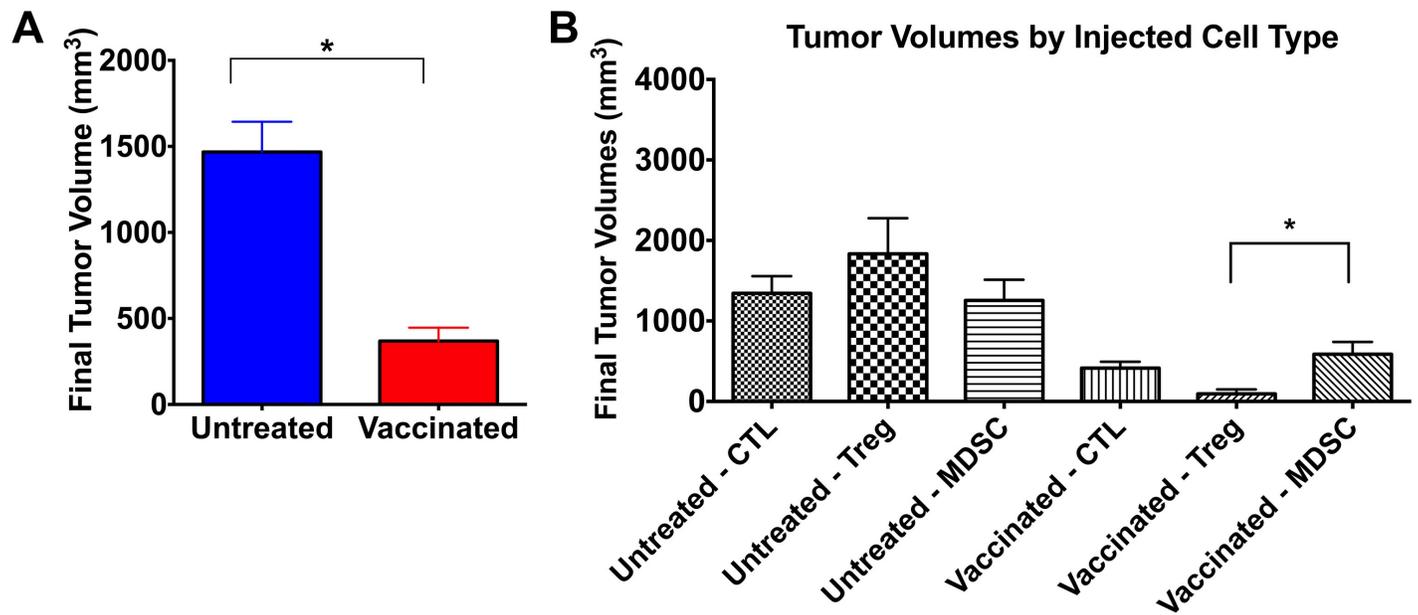

Figure 2

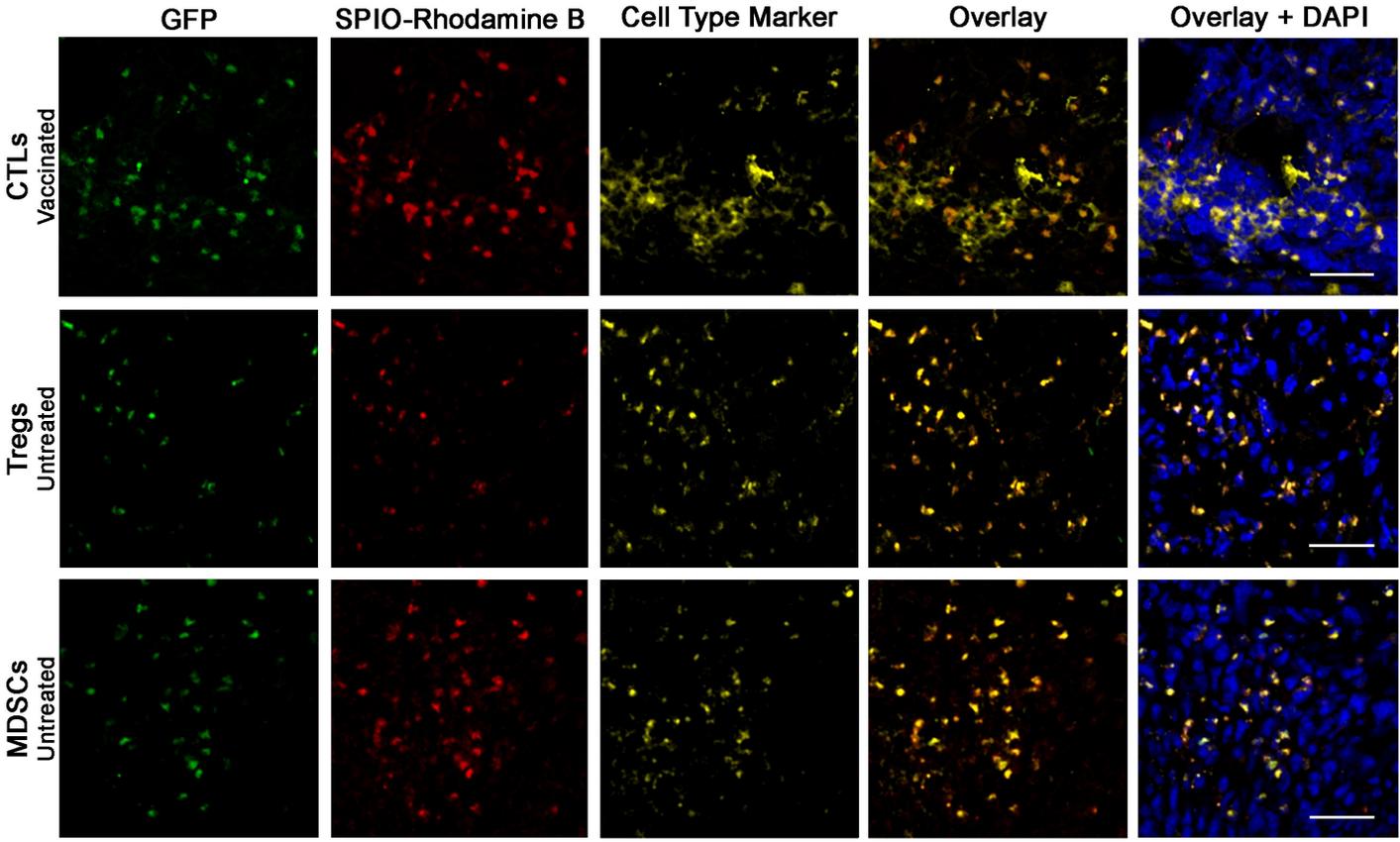

Figure 3

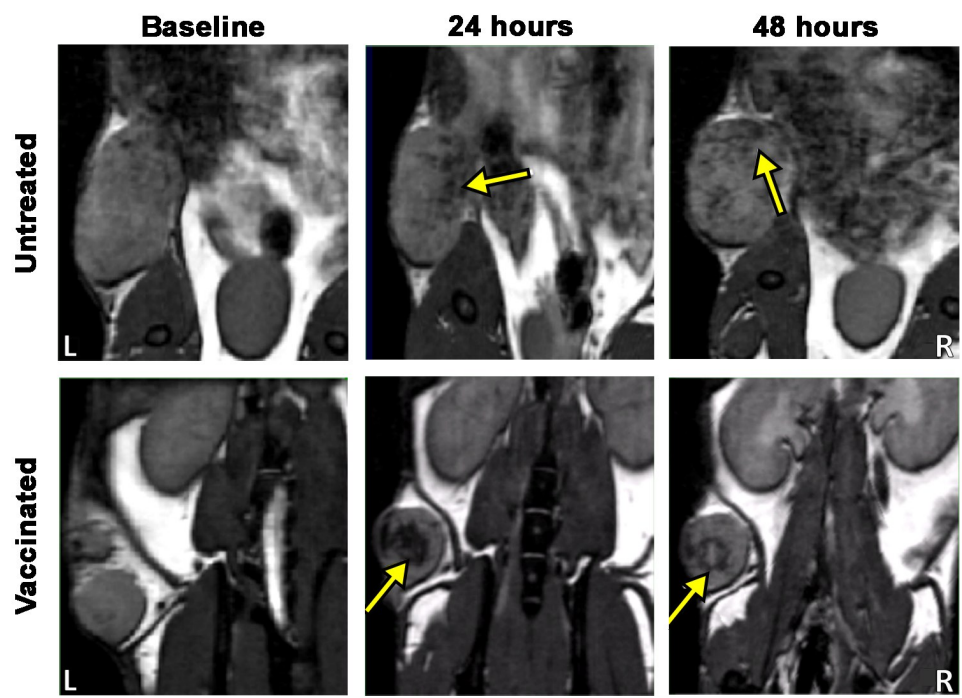

Figure 4

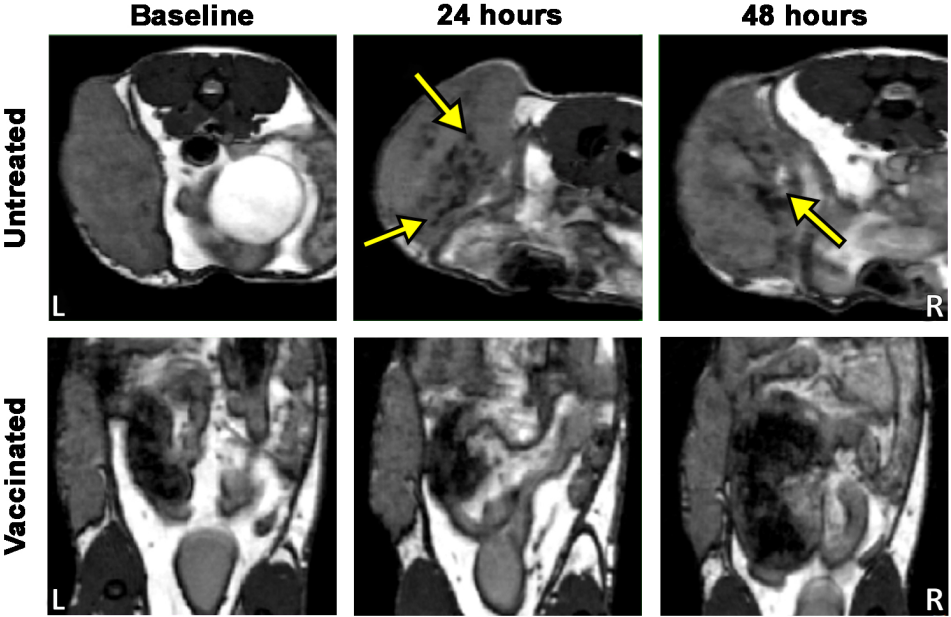

Figure 5

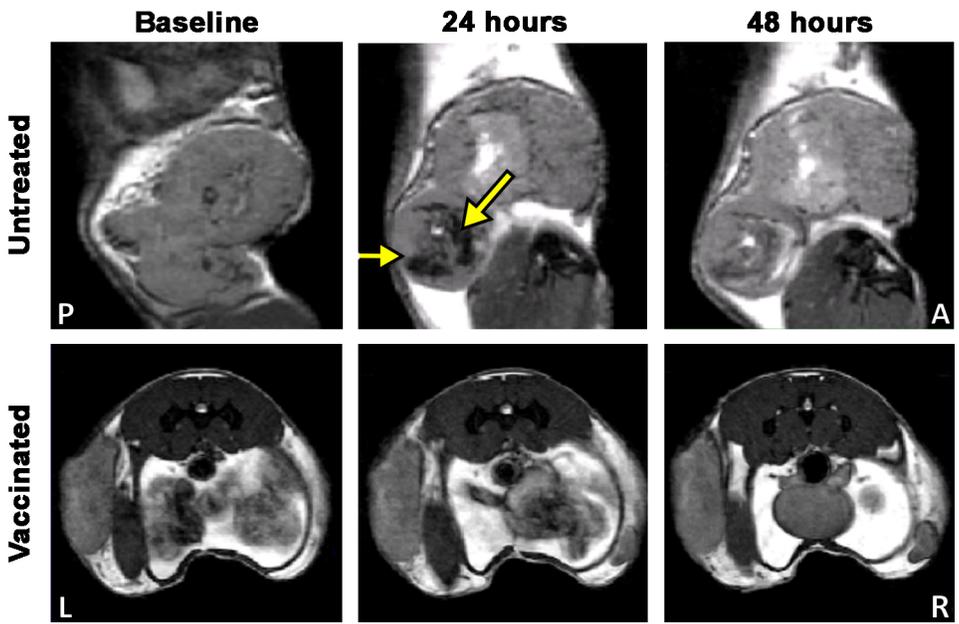

Figure 6

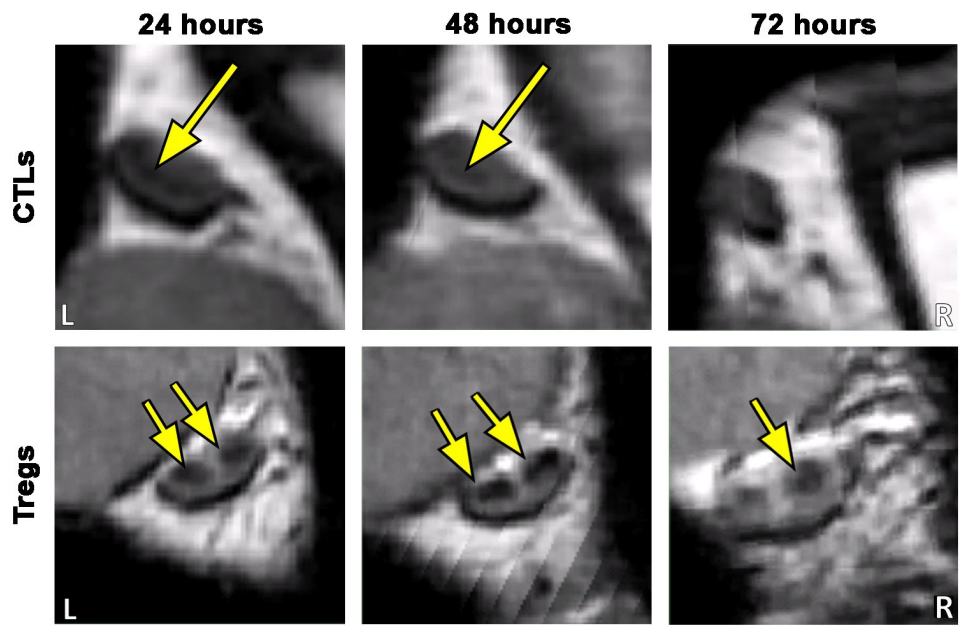

Figure 7

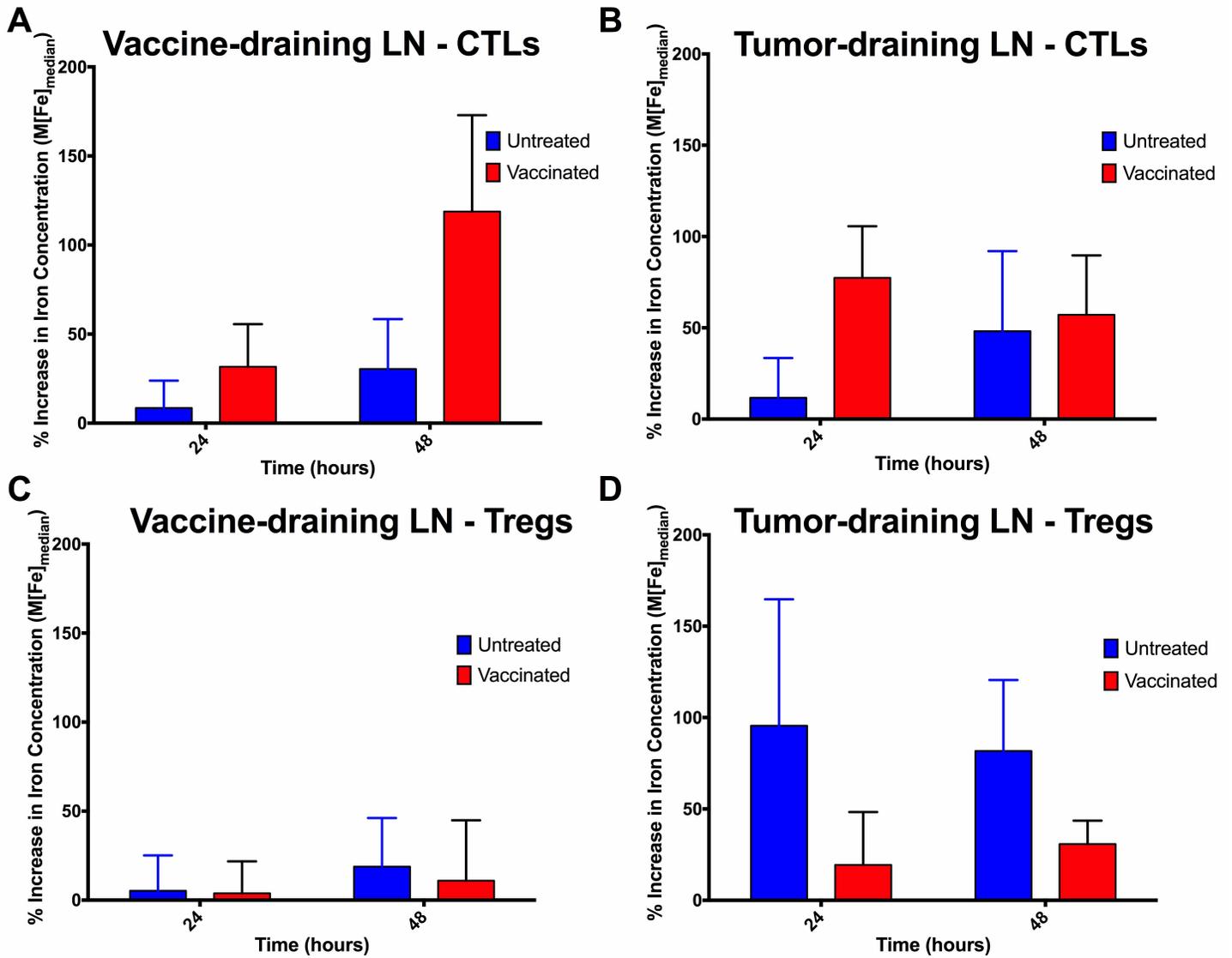

Figure 8

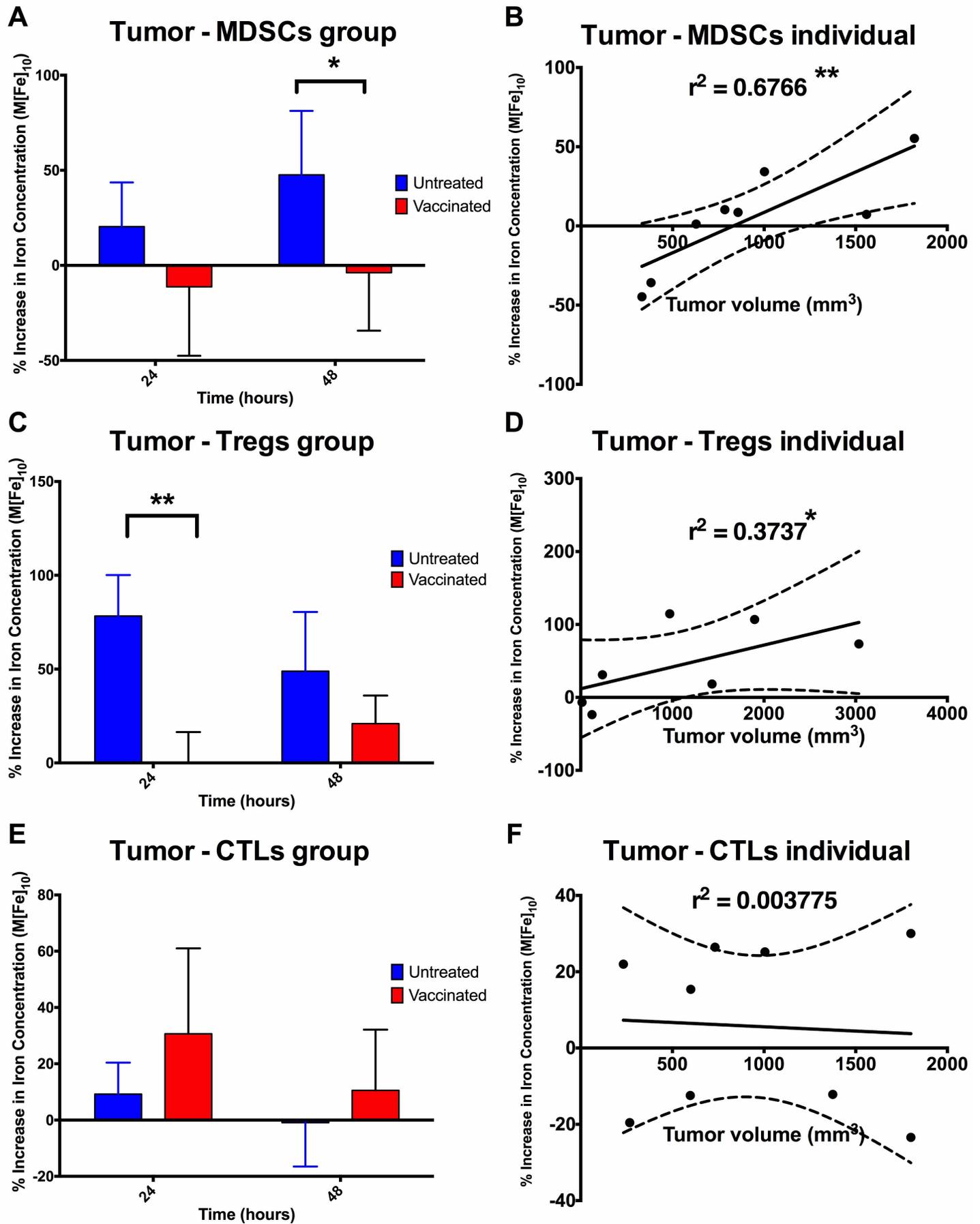

# Using MRI Cell Tracking to Monitor Immune Cell Recruitment in Response to a Peptide-Based Cancer Vaccine

Marie-Laurence Tremblay, Christa Davis, Chris V. Bowen, Olivia Stanley, Cathryn Parsons, Genevieve Weir, Mohan Karkada, Marianne M. Stanford, *Kimberly D. Brewer

## Supporting Methods

*Materials*. Culture media (IDEM and RPMI-1640), phosphate buffered saline (PBS), gentamycin, L-glutamine, 2-mercaptoethanol, penicillin-streptomycin (Pen-Strep; 10 000 U/mL), Hank's Buffered Salt Solution$^{++}$ (HBSS$^{++}$) were obtained from Gibco. Fetal calf (FCS) and fetal bovine (FBS) serums were obtained from Seradigm. Lipopolysaccharides (LPS), rapamycin, mitomycin-c, mouse IL-12, conc. hydrochloric acid (HCl), $K_4Fe(CN)_6$, paraformaldehyde (PFA), HEPES buffer, and bovine serum albumin (BSA) were obtained from Sigma. 10X permeabilizing buffer was obtained from eBioscience. SPIO nanoparticles (Rhodamine B Molday ION) were purchased from Molday, Biopal. C3 media was made by supplementing IMDM with 10% heat-inactivated fetal calf serum, 2 mM L-glutamine, 50 mM 2-mercaptoethanol, 100 U/ml penicillin and 100 μg/ml streptomycin). Complete RPMI media was made by supplementing RPMI-1640 with 10% FBS, 50 mM 2-mercaptoethanol, 100 U/mL penicillin and 100 μg/ml streptomycin.

All cell counts were performed using a hemocytometer and a cell dilution of 1:10 in trypan blue.

**Supporting Table S1** - *Primary Antibodies and Sources.*

| Target | Antibody Species | Clone | Source | Use | Dilution | Fluorescent Tag |
|---|---|---|---|---|---|---|
| CD4 | Rat anti-mouse | GK1.5 | eBioscience | CTL enrichment | 25 µg/mL | none |
| IgG1κ | Rat anti-mouse | 187.1 | BD Pharmingen/ BD Bioscience | CTL enrichment | 50 µg/mL | none |
| IgG | Goat anti-rat | Poly4054 | BioLegend | CTL enrichment | 10 µg/mL | none |
| CD28 | Hamster anti-mouse | 37.51 | BD Bioscience | CTL culture | 1 µg/mL | none |
| CD3 | Hamster anti-mouse | 145-2C11 | BD Bioscience | CTL plate coating | 2 µg/mL | none |
| CD8α | Rat anti-mouse | 53-6.7 | eBioscience | CTL labelling | 1:200 | efluor660 |
| CD3 | Rat anti-mouse | 17A2 | eBioscience | CTL labelling | 1:200 | PerCP |
| CD4 | Rat anti-mouse | GK1.5 | eBioscience | CTL labelling | 1:200 | PE |
| CD11c | Rat anti-mouse | N418 | eBioscience | CTL labelling | 1:200 | PE |
| CD25 | Rat anti-mouse | PC61.5 | eBioscience | Treg labelling | 1:200 | PE |
| FoxP3 | Rat anti-mouse | FJK-16s | eBioscience | Treg labelling | 1:200 | APC |
| CD4 | Rat anti-mouse | GK1.5 | eBioscience | Treg labelling | 1:200 | PerCP |
| CD11b | Rat anti-mouse | M1/70 | eBioscience | MDSC labelling | 1:200 | PE |
| Ly-6G (Gr-1) | Rat anti-mouse | RB6-8C5 | eBioscience | MDSC labelling | 1:200 | efluor660 |

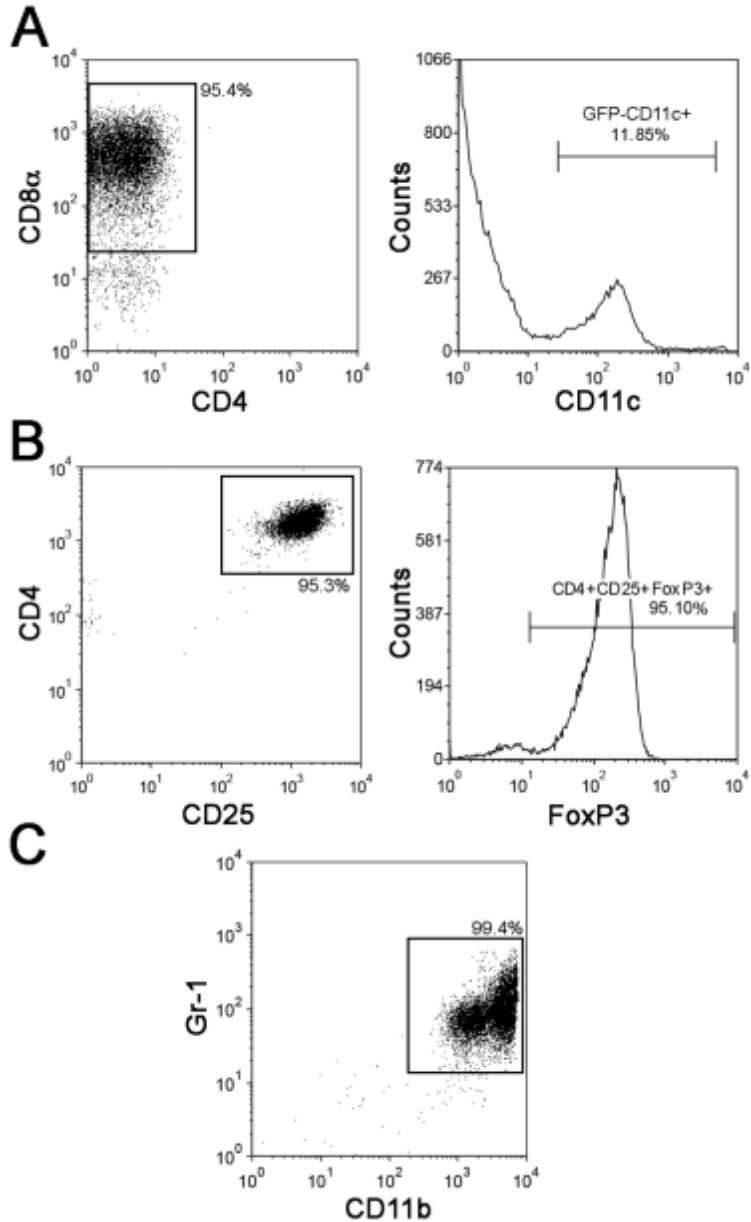

**Supporting Figure S1**. *Cell Purity of Isolated Cells.* Flow cytometry culture purity analysis for CTLs (A), Tregs (B) and MDSCs (C). Analysis was done immediately prior to iron labelling. Cells were gated for live cells and are representative of the full cell injection. CD11c control demonstrates that a small proportion (~10%) are APCs due to addition of APCs to culture for *in vitro* stimulation of T cells. FoxP3$^+$ cells were first gated for CD4$^+$CD25$^+$.

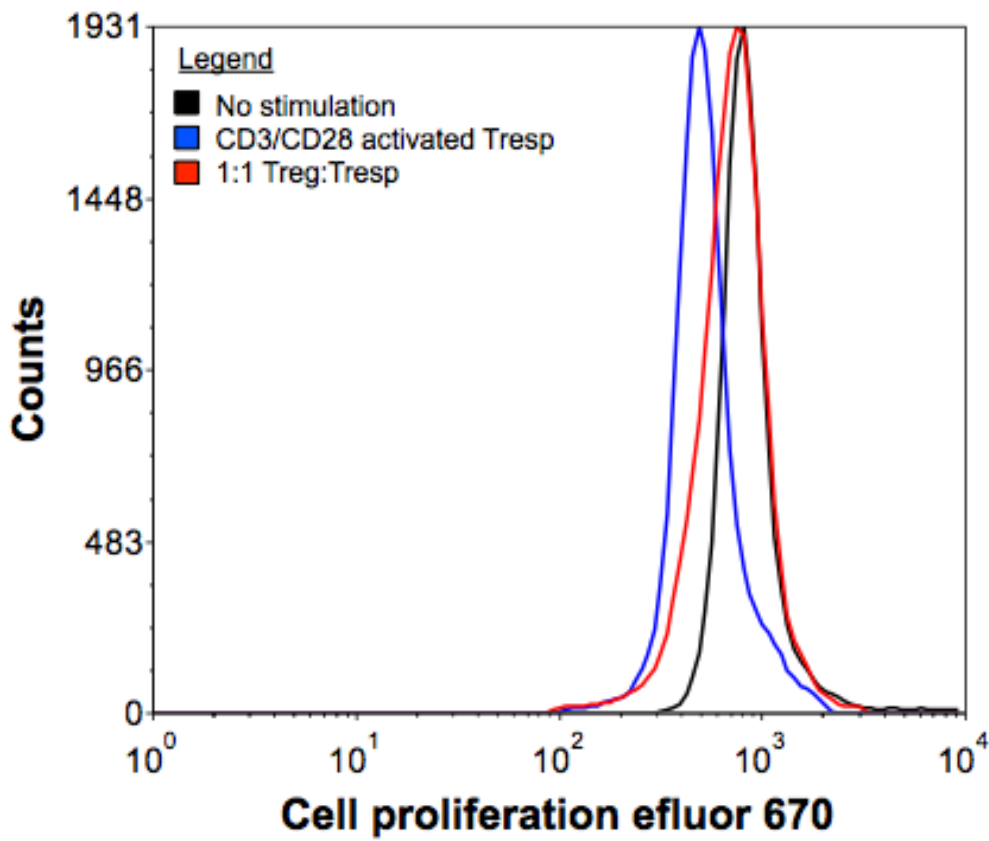

**Supporting Figure S2** - *Treg Supression Assay.* Treg-mediated suppression was measured using the cell proliferation dye efluor670 (eBioscience) at 0.5 μM. $CD4^+$ T cells (Tresp) were isolated from tumour-bearing C57BL/6 mice, incubated with efluor670, and activated using CD28 antibody on a plated $CD3^+$ 96-well plate. Tregs were isolated from transgenic $GFP^+$ mice. Tresps were cultured either alone (blue) or in the presence of Tregs at a 1:1 ratio (red). Cells were cultured for 72 hours and proliferation was determined by flow cytometry analysis gating for $GFP^-$ efluo670$^+$ cells. Tresp stimulation was confirmed by comparison to Tresp fixed at 0h of incubation (black).